\documentclass{kapproc} 

\usepackage{procps} 

\usepackage[dvips]{graphicx}

\upperandlowercase

\kluwerbib 

\newcommand{\xmm}{{\it XMM-Newton}}
\newcommand{\chandra}{{\it Chandra}}
\newcommand{\CXO}{{\it Chandra X-Ray Observatory}}

\begin{document}

\articletitle[X-ray Observations of DLS Clusters]
{X-Ray Observations of the Most Massive\\
 DLS Shear-Selected Galaxy Clusters}

\author{John P. Hughes,\altaffilmark{1}
 Ian Dell'Antonio,\altaffilmark{2} 
 Joseph Hennawi,\altaffilmark{3}
 Vera Margoniner,\altaffilmark{4}
 Sandor Molnar,\altaffilmark{1}
 Dara Norman,\altaffilmark{5}
 David Spergel,\altaffilmark{3}
 J. Anthony Tyson,\altaffilmark{4}
 Gillian Wilson,\altaffilmark{6}
and David Wittman\altaffilmark{4}
}

\affil{
\altaffilmark{1}Rutgers University, \
\altaffilmark{2}Brown University, \
\altaffilmark{3}Princeton University, \
\altaffilmark{4}Bell Labs, Lucent Technologies, \ 
\altaffilmark{5}CTIO/NSF AAPF, \
\altaffilmark{6}SIRTF Science Center
}


\anxx{John P. Hughes}

\begin{abstract}
We report on preliminary results of our X-ray survey of the most
massive clusters currently identified from the Deep Lens Survey (DLS).
The DLS cluster sample is selected based on weak lensing shear, which
makes it possible for the first time to study clusters in a
baryon-independent way.  In this article we present X-ray properties
of a subset of the shear-selected cluster sample.
\end{abstract}


\section{Introduction}

Nonbaryonic dark matter is apparently the dominant component of galaxy
clusters, yet all large samples of clusters to date are selected on
the basis of emission from the trace baryons they contain: visible
light from galaxies or X-rays from hot intracluster gas.  Now, for the
first time, we have a direct survey of mass in the Universe that is
unbiased with respect to baryons, the Deep Lens Survey (DLS). The DLS
is a deep, wide area, multicolor (BVRz') imaging survey being carried
out at the NOAO 4-m telescopes.  The survey was designed to detect
large scale structures in the Universe through weak lensing shear,
i.e., distortions to the shapes of distant background galaxies caused
by gravitational lensing of massive foreground objects.  The DLS team
has already shown that shear-selection is effective at finding new
galaxy clusters: Wittman et al.~(2001) report the discovery of CL
J2346+0045, the first galaxy cluster identified by its gravitational
effect rather than its radiation.

For this project, 12 square degrees of the DLS data (the maximum sky
area available at the time) were processed through the weak lens shear
pipeline (Wittman et al.~2003), revealing mass concentrations over a
wide range of redshifts.  These mass clusters were rank-ordered by
their shear signal and the top candidates were proposed for
observation by the \CXO\ in cycle 4.  Seven targets were awarded; an
additional candidate was available through the \chandra\ archive. One
of the principle goals of the follow-up X-ray observations is to
confirm that the DLS shear-selected clusters are associated with true
virialized, collapsed structures.  The basic X-ray information
(luminosity, size, morphology, extent of central concentration, and
gas temperature) obtained on the clusters will allow us to assess the
effect of shear-selection on the $L_X$--$T_X$ relation, the cluster
temperature function, and the relation of these to cluster mass.  The
full DLS X-ray cluster sample will also allow us to quantify the
false-positive rate of ``aligned filaments,'' i.e., line-of-sight
projections that appear as spurious mass concentrations in weak
lensing shear maps (e.g., White, van Waerbeke, \& Mackey 2002).

\begin{figure}[h]
\begin{center}
\hbox{
\resizebox{2.5in}{!}{\includegraphics{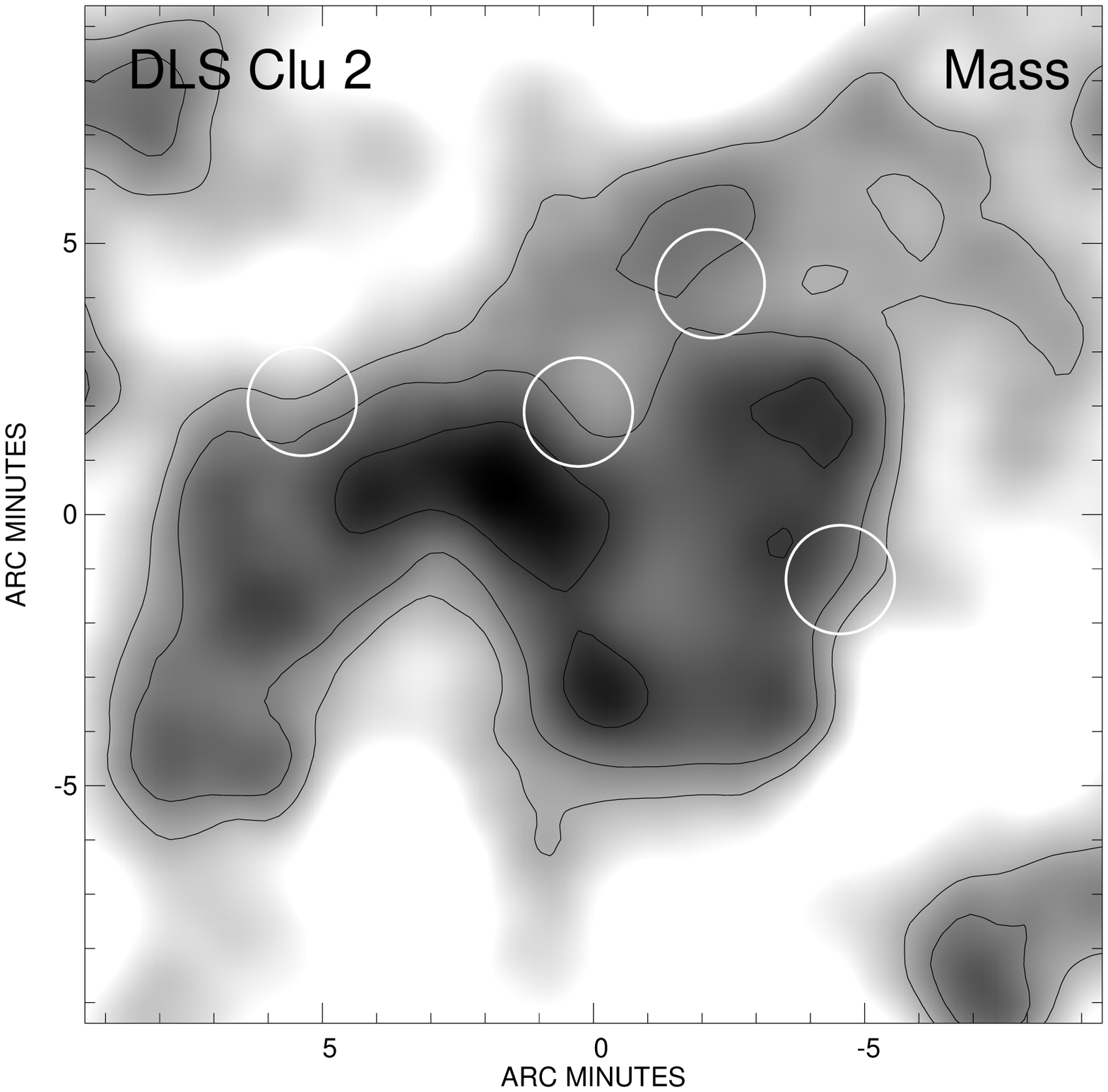}}
\resizebox{2.5in}{!}{\includegraphics{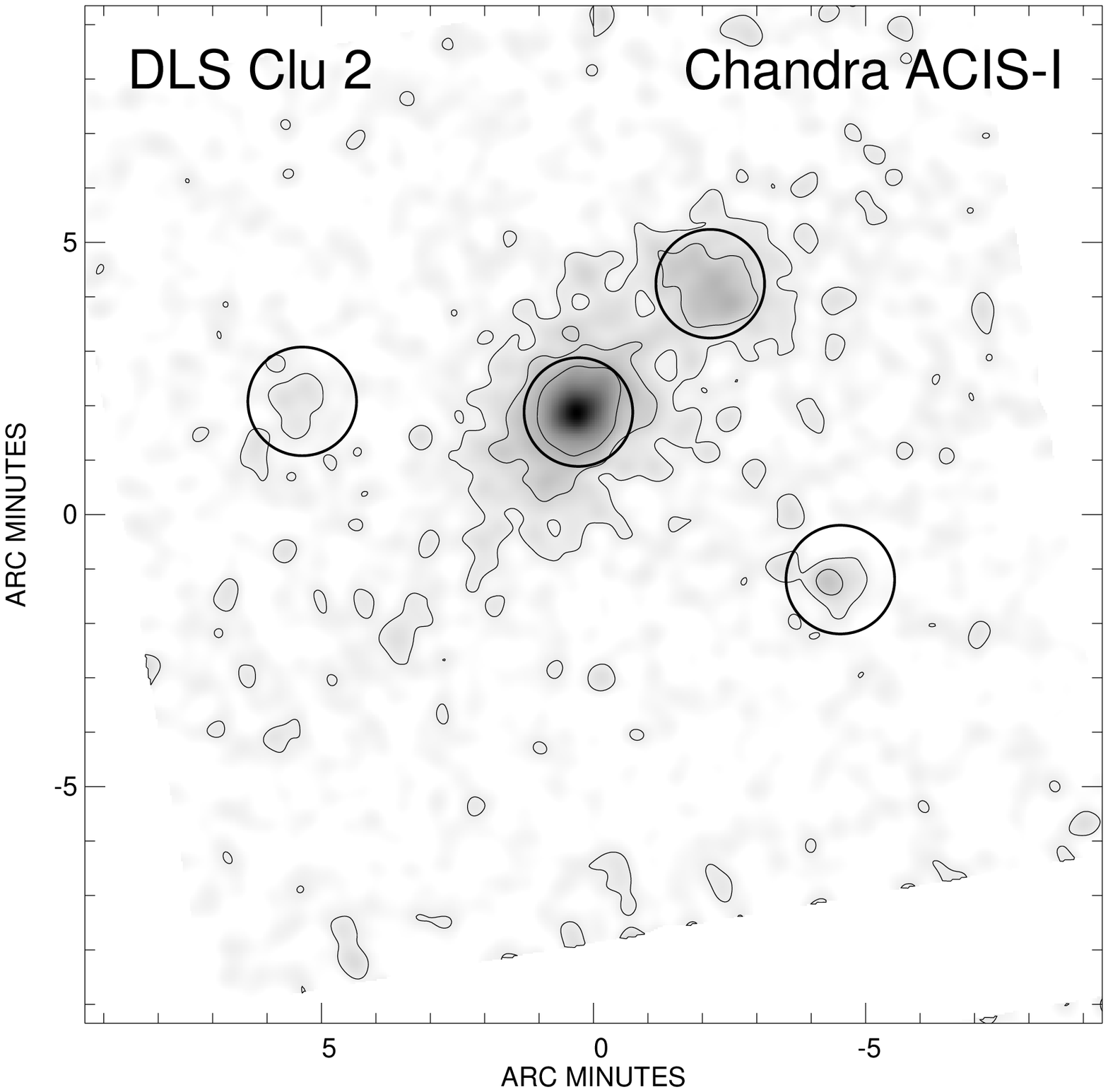}}
}
\vskip-0.17in
\hbox{
\resizebox{2.5in}{!}{\includegraphics{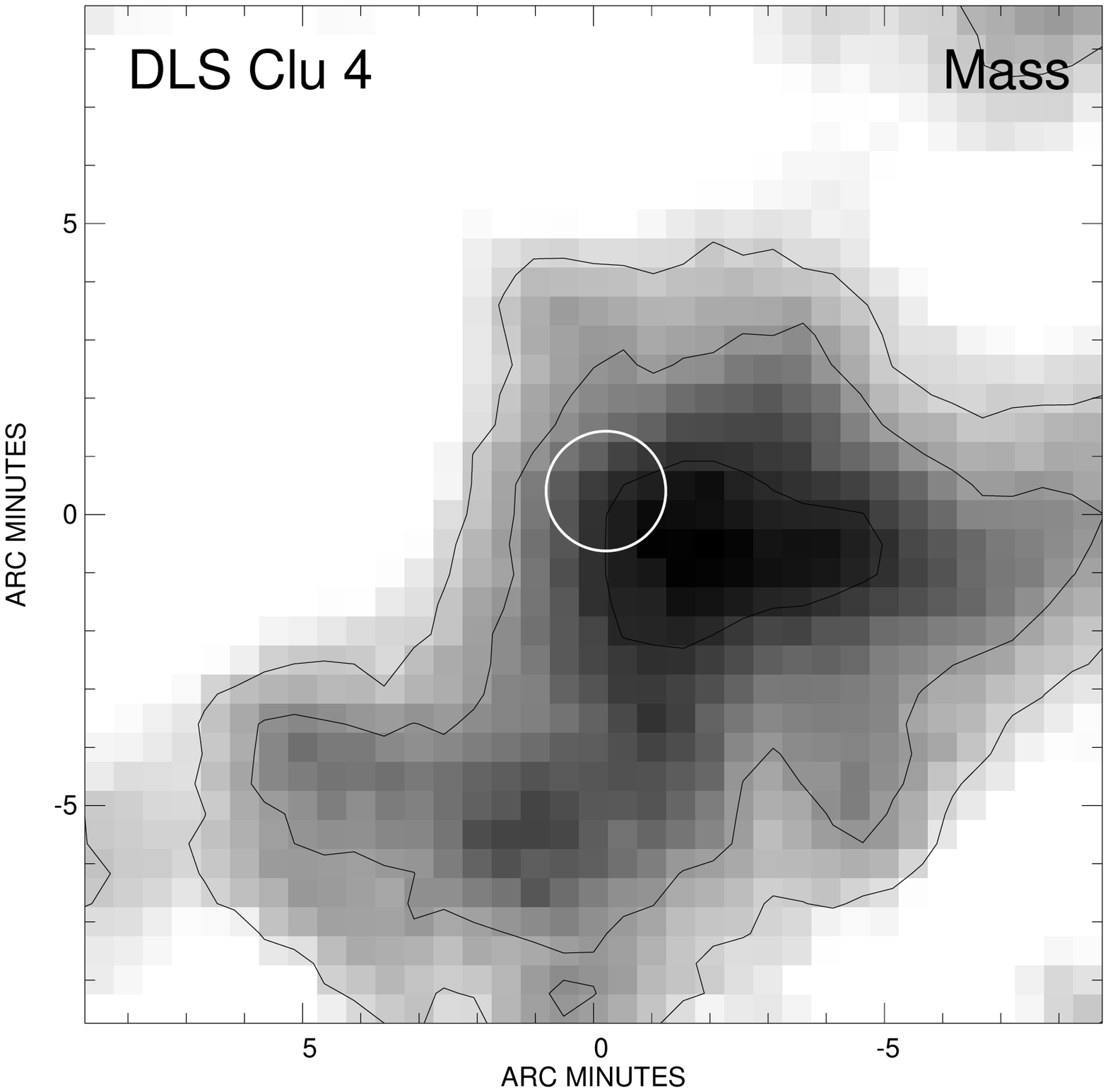}}
\resizebox{2.5in}{!}{\includegraphics{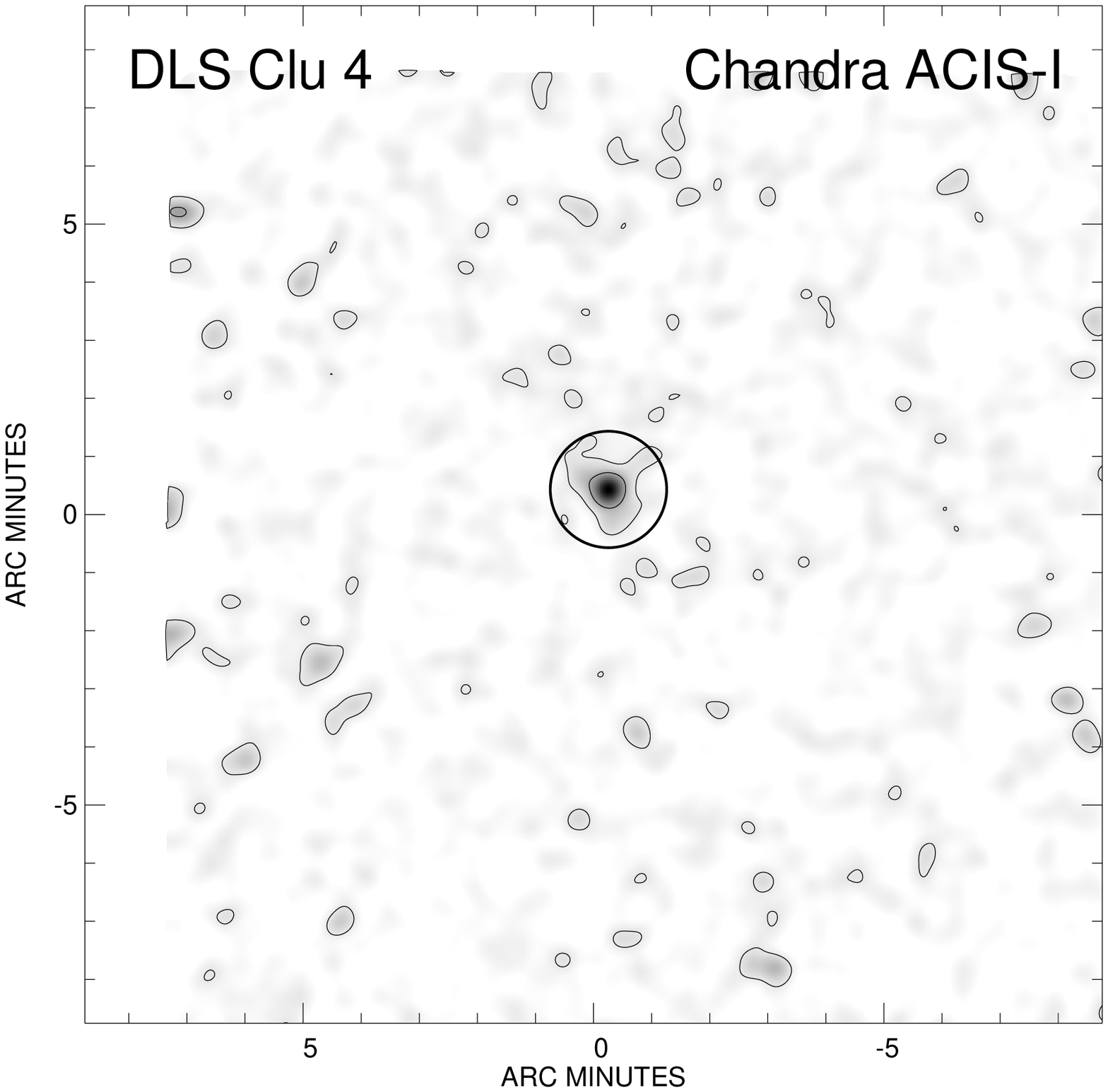}}
}
\vskip-0.17in
\hbox{
\resizebox{2.5in}{!}{\includegraphics{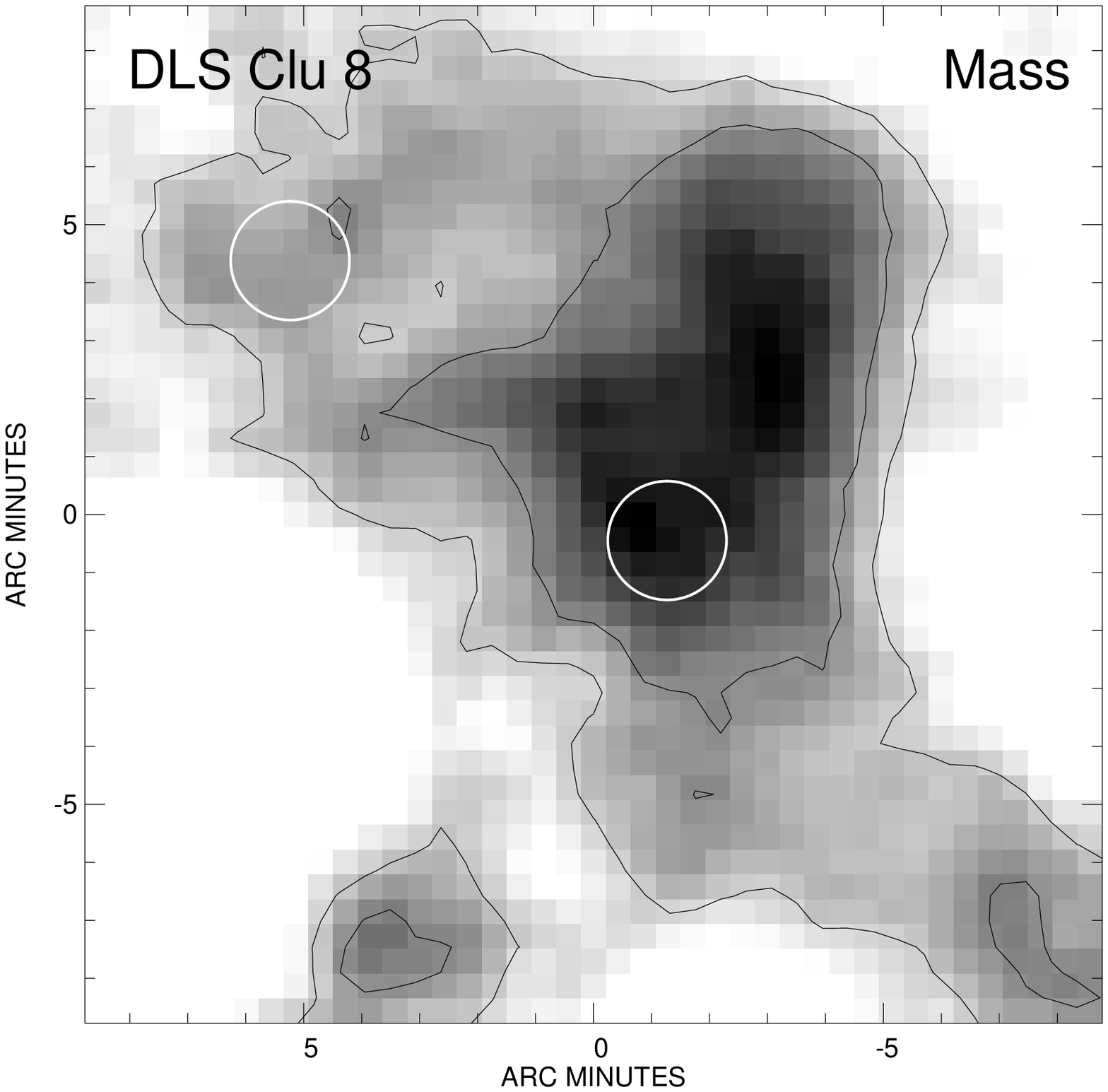}}
\resizebox{2.5in}{!}{\includegraphics{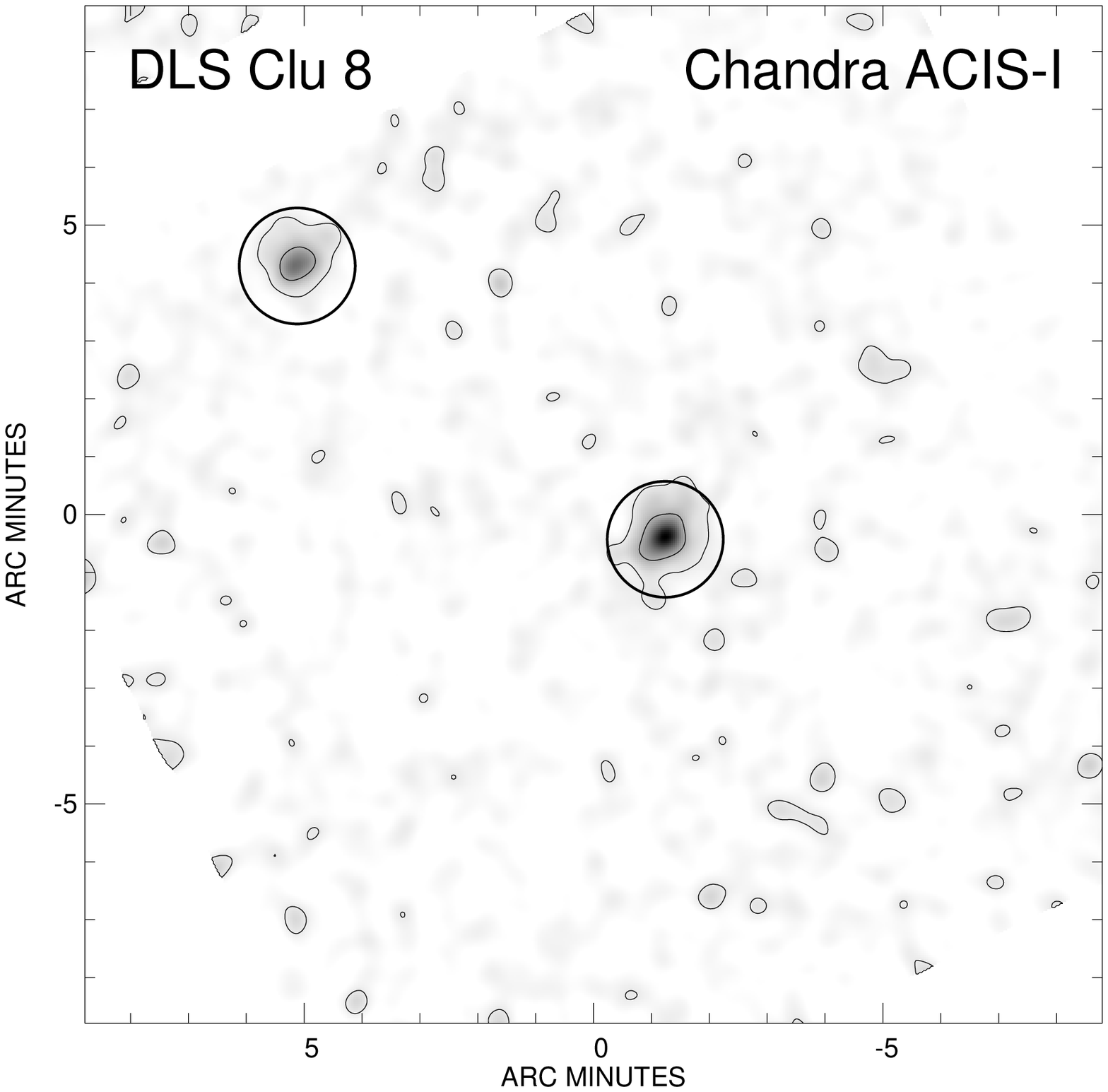}}
}
\end{center}
\vskip-0.47in
\caption{Projected mass (left) and X-ray surface brightness (right) 
of DLS mass clusters no.~2 (top), 4 (middle), and 8 (bottom).}
\end{figure}

\section{X-ray Observations}

We have confidently detected extended X-ray emission from at least
five of the eight DLS clusters in the \chandra\ cycle 4 sample.
Although the other three targets have been observed, our analysis is
not yet complete and we do not comment on them further here.

Figure 1 shows maps of the projected mass (left panels) and X-ray
surface brightness (right panels) over the 0.5--2 keV band for several
of the DLS mass clusters. In the X-ray images serendipitous 
point sources have been removed and circles denote the locations of
extended X-ray sources detected at signal-to-noise ratios greater than
3. The two highest-ranked shear-selected clusters are strong X-ray
sources, with multiple subclusters associated with each system. This
can be seen clearly in the X-ray image of DLS cluster 2 (top panel of
Figure 1).  In the optical images there are a large number of bright
galaxies that are presumably cluster members, however no published
redshifts are available.  The fourth ranked cluster (middle panels) is
unusual in having only a single X-ray component.  It is centered on a
galaxy at a redshift of $z=0.19$. The last ranked cluster in the
\chandra\ cycle 4 sample (bottom panels) shows two significant
extended X-ray sources. The southwestern component was confirmed by
the DLS team as a massive cluster at $z=0.68$ (Wittman et al.~2003)
and even shows a giant arc from strong lensing.  There is no redshift
available for the northern X-ray concentration.

\begin{table}[ht]
\caption{X-ray Properties of DLS Mass Clusters.}
\begin{center}
\begin{tabular}{|ccccc|l}\hline\hline
   &     &  $F_X$ (0.5--2 keV)       & $L_{\rm bol}$  & 
 $M_{200}$  \vrule height 10pt width 0pt  &\cr
 Cluster & $z$ & (erg s$^{-1}$ cm$^{-2}$) & (erg s$^{-1}$) 
 & ($10^{14}\,M_\odot$)  \vrule height 4pt width 0pt depth 4pt & \cr \hline
DLS Clu 1  
 & 0.298 & $6.7\times10^{-13}$ 
 & $8.5\times10^{44}$ & 5.7\vrule height 10pt width 0pt & \cr
DLS Clu 2  
 & $\sim$0.2 & $2.9\times10^{-13}$ 
 & $1.1\times10^{44}$ & 1.9 & \cr
DLS Clu 4 
 & 0.1894    & $1.4\times10^{-14}$
 & $4.3\times10^{42}$ &  0.4 &\cr
DLS Clu 7 
 & $\sim$0.4 & $1.9\times10^{-14}$
 & $3.1\times10^{43}$ &  1.0 &\cr
DLS Clu 8 
 & 0.68      & $2.4\times10^{-14}$
 & $1.7\times10^{44}$ &  2.4\vrule height 0pt width 0pt depth 4pt
  &\cr \hline\hline
\end{tabular}
\end{center}
\vskip-0.15in
\end{table}

In Table 1 we present selected numerical results for the five X-ray
clusters associated with shear peaks.  In each case only values for
the X-ray cluster component with the highest flux are given.  Redshift
information for clusters 2 and 7 is not yet available; we give very
preliminary luminosity and mass estimates based on approximate
redshifts from the magnitudes of the member galaxies. Photometric
redshifts for all X-ray clusters are in the process of being determined 
from our imaging data.
%
%
We used the X-ray luminosity-temperature relation (Arnaud \& Evrard
1999) to estimate cluster temperatures and then the mass-temperature
relation (Evrard, Metzler, \& Navarro 1996) to determine the mass
within a density contrast of $\delta_c=200$.  These relations are
derived from or calibrated against low-redshift clusters; at this
point we have not made any adjustments for the redshift range of our
sample. 
Still the mass estimates are consistent with simulations by our group
that show an expected mass range for DLS shear-selected clusters
extending from roughly $5\times 10^{13}\, M_\odot$ to $10^{15}\,
M_\odot$ with a peak near $3\times 10^{14}\, M_\odot$.  An additional
direct measurement of the mass can be made from the shear maps; this
work is in progress.
%

%

\section{Conclusions}

The DLS is clearly discovering true three-dimensional clusters of
mass, hot X-ray gas, and galaxies. We find a wide range of X-ray
luminosity for clusters with similar weak lensing shear. Nearly every
shear peak contains multiple X-ray clusters, while the high mass
clusters we detect are particularly complex with up to 4 or 5
individually resolved subcomponents. Work in progress on this unique
galaxy cluster sample includes measuring X-ray temperatures from
\chandra\ and \xmm\ spectra, deriving redshifts and velocity
dispersions from ground-based spectroscopy, more detailed mass
determinations, and additional numerical simulations.


%

\begin{acknowledgments}
This work was partially supported by \chandra\ grant GO3-4173A and NASA
LTSA grant NAG5-3432 to JPH.

\end{acknowledgments}


\begin{chapthebibliography}{1}
\bibitem{arnaud}
Arnaud, M, \& Evrard, AE~1999, MNRAS, 305, 631 

\bibitem{evrard}
Evrard, AE, Metzler, CA, \& Navarro, JF~1996, ApJ, 469, 494

\bibitem{white}
White, M, van Waerbeke, L, \& Mackey J~2002, ApJ, 575, 640


\bibitem{wittman01}
Wittman, D., Tyson, JA, Margoniner, VE, Cohen, JG, \& Dell'Antonio, IP~2001, 
ApJ, 557, L89

\bibitem{wittman01}
Wittman, D., Margoniner, VE,  Tyson, JA, Cohen, JG, \& Dell'Antonio, IP~2003, 
ApJ, submitted

\end{chapthebibliography}
\end{document}